\documentstyle[preprint,prl,aps]{revtex}
\begin{document}
\draft

\begin{title}
{\bf Structural phase transformations via first--principles simulation}
\end{title}

\author{P. Focher, G.L. Chiarotti, M. Bernasconi, E. Tosatti}
\address{
International School for Advanced Studies (SISSA),
Via Beirut 2, I-34014 Trieste, Italy
}
\author{M. Parrinello}
\address{
IBM Research Division, Zurich Research Laboratory, CH-8803 R\"uschlikon,
Switzerland}
\date{\today}
\maketitle

\begin{abstract}
We present a new simulation scheme for structural phase
transitions via first principles molecular dynamics. The method is obtained by
combining the Car--Parrinello method for {\sl ab initio} simulation with
the Parrinello--Rahman method to account for variable cell shape.
We demonstrate the validity of our approach by simulating
the spontaneous transformation of silicon from diamond to simple
hexagonal phase under high pressure.
\end{abstract}


\narrowtext

Solid--solid structural phase transitions with
increasing  of pressure and/or
temperature are ubiquitous in condensed matter.
Existing {\sl ab--initio} theory of these transitions has been
mostly restricted to the study of relative energies of
known structural phases at different volumes and at zero temperature.
Despite a remarkable level of understanding and accuracy attained in
this area, mostly  through  Local Density Approximation (LDA) based
calculations \cite{pippo1}, the main goal of simulating these
transitions under real conditions of temperature and pressure, while
still retaining the first principles accuracy required for the
prediction of new structures, has not yet been  achieved.
In particular, heavily first order structural phase transitions do not
take place in {\sl ab--initio} Car--Parrinello (CP) molecular
dynamics simulation \cite{CP}, in spite of a correct description of
both energetics and thermal fluctuations. The main problem lies in the
(mandatory) use of periodic boundary conditions (pbc) within a rigid
simulation cell, which does not allow for the necessary density and shape
fluctuations.

In the early 80's Parrinello and Rahman (PR) proposed a
lagrangian which, by allowing for a variation of the pbc guided by the
internal stress, made simulations of solid--solid phase
transitions possible \cite{PR}. Applications
of this method since then have been numerous \cite{pippo2,PRrev}, and further
refinements also proposed \cite{pippo3}.
However, existing implementations of the PR scheme are based on
empirical -- rather than {\sl ab--initio} -- potentials: this
restriction reduces the predictive power, and thus the
general usefulness of the method.
A recent step forward was taken by Wentzcovitch {\sl et al.} \cite{W},
who proposed a method aimed at combining {\sl ab--initio}
Hellmann-Feynman forces with a PR like variable cell.
However the
method has been shown to work only as a tool for structural
parameters optimization.

In this letter, in the spirit of the original PR work, we propose a
lagrangian that, while describing the ionic dynamics
with first principles accuracy, still allows for
the symmetry breaking fluctuations necessary for phase transitions
to take place {\sl spontaneously} during the simulation.
We have applied our new
method to the study of pressure induced transformations in silicon,
in particular we have successfully simulated the real-time,
spontaneous diamond $\to$ simple
hexagonal transformation.

Following Parrinello and Rahman we introduce in the
CP lagrangian of Ref.~\cite{CP} nine
additional cell degrees of freedom,
represented by a matrix ${\bf h} = \{\bf a,\bf b,\bf c \}$, where
${\bf a,\bf b,\bf c}$ are the cell edges.
A generic point in real space will be expressed as ${\bf r}={\bf h}{\bf s}$,
where ${\bf s}$ are the so called scaled
variables whose components are the projections of ${\bf r}$ on the cell edges.
The new CP--PR combined lagrangian is:
\begin{eqnarray}
  {\cal L} &=& \mu \sum_i\int d{\bf s} \,|\dot \psi_i({\bf s})|^2 +
       \frac{1}{2}\sum_{I} M_I({\bf \dot S}_I^t{\bf{\cal G}}{\bf \dot S}_I)
               - E \left[ \left\{ \psi_i\right\} ,
        \left\{ {\bf h}{\bf S}_I \right\} \right] +              \nonumber \\
          & & \sum_{ij} \Lambda_{ij}
    \left( \int d{\bf s} \,\psi^*_i({\bf s}) \psi_j({\bf s}) -\delta_{ij}
      \right) + \frac{1}{2} W\mbox{Tr}({\bf \dot h}^t{\bf \dot h}) - p\Omega ,
                                                      \label{lagrCP}
\end{eqnarray}

Here $\psi({\bf s})$ are the Kohn--Sham fields represented
in term of scaled variables ${\bf s}$, ${\bf S}$ are the scaled
ionic coordinates, ${\bf{\cal G}}={\bf h}^t{\bf h}$, and
$E \left[ \left\{ \psi_i\right\} , \left\{ {\bf h}{\bf S}_I \right\} \right]$
is the LDA \cite{PZ} energy functional,
supplemented with direct ionic Coulomb repulsions. The fourth term
in Eq.~\ref{lagrCP} is the orthonormality constraints on the
$\psi_i$, $\Lambda_{ij}$ being lagrangian multipliers.
$\mu$ and $W$ are inertia parameters controlling the time scale of
electronic motion \cite{CPrev} and of ${\bf h}$ \cite{PRrev} respectively,
$M_I$ are the nuclear masses. Finally
$p$ denotes the externally applied hydrostatic pressure and
$\Omega=\det({\bf h})$ is the volume of the cell.

The $\psi({\bf s})$ represent ``stretched'' wave functions, invariant
with respect to variation of
${\bf h}$. They are related to the real wave functions $\psi_h({\bf r})$
by:
$
   \psi_h({\bf r}) = \Omega^{-1/2} \psi({\bf h}^{-1}{\bf r})
                = \Omega^{-1/2} \psi({\bf s}).
$
Note that $\psi_h({\bf r})$ depends on ${\bf h}$ due to the normalization
condition on the cell volume. Hence, the
$\psi_h({\bf r})$ are not  independent fields,
once ${\bf h}$ is itself present as lagrangian degrees of freedom.
For a rigid cell (${\bf \dot h} = 0$) Eq.~\ref{lagrCP}
reduces to the original CP lagrangian \cite{CP}, plus the constant
term $p \Omega$.

Lagrangian (\ref{lagrCP}) gives rise to the following equations of motion:
\begin{eqnarray}
 \mu \ddot{\psi}_i({\bf s})&=&-{{\delta E}\over {\delta \psi^*_i({\bf s})}}
     + \sum_{j} \Lambda_{ij} \psi_j({\bf s})        \label{CPeqmoto1}   \\
  \ddot{\bf S}_I^\alpha & = & -\frac{1}{M_I}
     {{\partial E}\over{\partial {{\bf R}_I^\beta}}}
      ({\bf h}^{t})^{-1}_{\beta\alpha} -
      {\bf{\cal G}}^{-1}_{\alpha\beta}{\bf\dot{\cal G}}_{\beta\gamma}
      {\bf \dot S}_I^\gamma ,                     \label{CPeqmoto2}   \\
  \ddot{\bf h}_{\alpha\beta} & = & \frac{1}{W} \left({\bf\Pi}_{\alpha\gamma} -
      p\delta_{\alpha\gamma} \right) \Omega({\bf h}^{t})^{-1}_{\gamma\beta} ,
                                           \label{CPeqmoto3}
\end{eqnarray}

with
\begin{eqnarray}
  {\bf\Pi}_{\alpha\gamma}= \frac{1}{\Omega}\left( \sum_I M_I
   \left( {\bf \dot S}_I^t{\bf{\cal G}}
          {\bf \dot S}_I \right)_{\alpha\gamma}      -
    {{\partial E}\over{\partial{\bf h}_{\alpha\delta}}}
      {\bf h}^t_{\delta\gamma}\right).
\label{stress}
\end{eqnarray}
Where ${\bf R}={\bf h}{\bf S}$ and
greek letters indicate the components of
vectors and matrices, with implicit sum over
repeated indices.
Due to our normalization choice for the $\psi({\bf s})$ the equations
of motion for the electronic wave--functions (Eq. \ref{CPeqmoto1}) are
formally the same of the original CP ones \cite{CP}. Eq. \ref{CPeqmoto2} is
also the same as in classical PR, with the classical forces replaced by
the quantum--mechanical ones \cite{CPrev}. Eq. \ref{CPeqmoto3} are the
equations of motion for the additional cell degrees of freedom.
The forces acting on them are due to the
imbalance between the total microscopic stress tensor ${\bf\Pi}$ and the
external applied pressure $p$.
The first term in Eq. \ref{stress} is the thermal stress while the
remaining term is the quantum--mechanical internal stress
\cite{NM}.
The Eqs. \ref{CPeqmoto1}--\ref{CPeqmoto3} thus combine the fictitious
electronic dynamics as formulated by CP
with the PR fictitious cell dynamics, which allows the
sampling of the ionic isoentalpic--isobaric ensemble.

We have tested our method  by
simulating a pressure induced transition in Si.
Si has a quite complicated phase
diagram with many phases appearing at high pressure \cite{Duclos}.
The experimentally determined sequence up to 30 GPa is:
diamond (D) $\to$ $\beta$-Sn
$\to$ simple hexagonal (sh), with transition pressures at 11 and 13-16
GPa respectively.  The domain of stability of  $\beta$-Sn is therefore
very narrow in contrast with those of D and sh.
Other phases appear at considerably higher pressures \cite{Duclos}
and will not be investigated in the present work.
The transition pressures have been calculated with
standard zero temperature LDA total energy calculations
\cite{MC} which essentially
reproduce the experimental values. Both sh and $\beta$-Sn structures
show metallic character, unlike the covalent D structure.

To study this phase diagram we started from an fcc periodically
repeated cell containing 54 atoms, initially arranged in a diamond
lattice.
{\sl Ab initio} norm--conserving pseudopotential of the
Kleinman--Bylander \cite{KB} form were constructed from the
pseudopotential of Ref. \cite{BHS} retaining only s--nonlocality.
Kohn--Sham wave--functions were expanded in
plane waves up to a kinetic energy cutoff of 12 Ry.
The inertia parameters $\mu$ and $W$ were fixed to the values of
300 a.u. and 4.05 Si mass respectively. In particular
$W$ was tuned in order to
maximize energy exchange between the ionic degrees of freedom and the cell.
The integration step was $1.45\times 10^{-16}$ s.
The temperature of the ions was controlled by a Nos\'e thermostat
\cite{nose}, and when necessary the electrons were maintained close to their
ground state by means of the technique of Ref. \cite{BL}.
Full details of the calculation will be given elsewhere.

We did not allow for a variation of the plane waves number
during the simulation, as would be necessary to work at constant energy cutoff
in presence of volume fluctuations.
The number of plane waves was fixed
by a 12 Ry cutoff with our  initial cell and a lattice
constant of 5.36 \AA.
We restrict the k-sum to the $\Gamma$
point of the supercell Brillouin Zone. For our 54 atom cell this
gives an energy per particle $\sim$20 mRy/atom higher than the fully converged
value. This error, as well as that introduced by the constant plane
waves number are expected to affect only  the numerical values
of the transition pressures. Since
a large uncertainty on the phase boundaries is anyway likely, due to  the
hysteresis phenomena related to our small cell and short simulation time,
we have not tried to improve these aspects of the calculation.
All these limitations can be reduced  by increasing
the dimension of the cell and the energy cutoff.

We started the simulation by heating our
D-Si system  up to 300  $^\circ$K,
and equilibrating both ions and cell degrees of freedom for
$\sim 1$ ps.
Skipping a possible search of the $\beta$-Sn phase, of unlikely value
in view of our relatively limited accuracy and of the narrow range of
stability, we  raised the pressure from zero to a large value of 30
GPa. This was done in steps, but without waiting for equilibration at
each step, in an overall time of $\sim$0.3 ps.

Initially the system followed the equation of state
of the D structure (within our approximations) up
to a pressure of 30 GPa.
The cell edges and their relative angles oscillate isotropically for a while
as a function
of time (fig. \ref{1}), with
large volume fluctuations  due to the compressive shock, following
the sudden pressure increase.
A dramatic event spontaneously
took place 0.6 ps after compression, the cell
undergoing an evident change of shape.
 At this point we
further raised the pressure up to
42 GPa to allow for a faster stabilization of the novel crystal
structure. We then let the system equilibrate without any further
perturbations for $\sim$5.5 ps. The change in shape of cell
was accompanied by radical changes in the electronic and ionic structure.
In fig. \ref{1}c we show as a function of time
two peaks in the structure factors which are characteristic of the D
and sh structures. It is clearly seen that the initial D structure has changed
into a new one, which we will show to be simple hexagonal.
The D$\to$sh transformation occurs via  large, concerted, atomic motions.
A simple transformation path has not so far been further rationalized.
The crystal structure of the new phase is first of all
 reflected
in the average
pair correlation function  $g(r)$
and angle distribution function $f(\theta)$
reported in fig. \ref{2}, both of which are only compatible with a sh
structure, with
$f(\theta)$  peaks at 60, 90, 120 and 180 degrees.
 However the peaks of both  $g(r)$
and  $f(\theta)$  are too broad for a perfect
crystal structure.  Visual inspection reveals
 defected hexagonal planes. Interference
between these planes gives rise to the strong Bragg reflection
in fig. \ref{1}c.  These planes are however
somewhat misaligned which explains the large
width of the peak in  $f(\theta)$ at 90 degrees. The
defect can be seen as a missing
3--atoms segment in one of the hexagonal rows.
Thus addition of three atoms should remove the defect.
To prove this we have constructed an ``ideal'' structure compatible with
the shape of the MD cell.
The primitive cell in reciprocal space of this ideal crystal
is defined
using three linearly independent
$\bf k$ vectors corresponding to zero order Bragg reflections.
This construction gives rise to
a triclinically distorted sh (9 \% in plane distorsion and
a misalignement of about two degrees between adiacent planes containing
the hexagons) made of 57 atoms in the original MD cell.
With a further relaxation of the cell the 57 atoms crystal
trasforms in a perfect sh with $a= 2.52 {\rm \AA}$
and $c= 2.26 {\rm \AA}$~\cite{nota}.
We have separately
calculated the properties of this perfect sh and its energy
at V=12.3 \AA$^3$/atom is -7.866 Ry/atom.
This 57-atoms $\Gamma$-point energy is close to the exact
value at full k-point convergence (-7.861 Ry/atom).
This energy value is also close
to that of the defected 54 atoms cell (-7.863 Ry/atom at V=12.2 \AA$^3$/atom).
It is remarkable how
little is the energy cost of the defect.
In Fig. \ref{3} we report one of the
hexagonal planes. The defected part of the structure is clearly visible there.

Electronic states also undergo
a transition together with the D$\to$sh structural transformation.
Strikingly (fig. \ref{4}) both
the starting D configuration {\sl and}
 the final defected sh  show a gap at the Fermi
level (${\rm E_F}$).
This is at first
surprising, since the perfect sh crystal is a metal
 with only a pseudogap just below
${\rm E_F}$ \cite{MC}. This is clearly visible in fig. \ref{4} where
we report the electronic density of states for the sh perfect crystal,
along with the  $\Gamma$-point electronic states
of the cells containing 57 atoms arranged in the triclinically distorted
and perfect sh structures.
As seen there, the defect in the 54 atoms cell
deforms the cell in such a way to enhance the pseudogap and
to cause  ${\rm E_F}$ to drop
to the bottom of the pseudogap, in a region of very low density of states.
Therefore
the excellent stability of the 3-atoms defect (plus deformation)
results from a kind of
charge compensation  -- the defect is an acceptor and getters away the
excess of electrons populating the high energy states.

At this point, having obtained the D-sh conversion,
 it was natural to ask whether the reverse sh$\to$D
transformation could also be observed upon a decompression procedure.
We therefore proceeded in reverse by decreasing the pressure back to zero.
Wary of difficulties and metastable phases encountered by experimentalist
along this path, we tried a very careful and slow
 decompression process, allowing the system to equilibrate at
every intermediate step. This took $\sim$4 ps
with $p$ decreasing from 42 GPa to -12 GPa.
In this pressure range (still with T= 300
$^\circ$K) very little happened to the defected sh structure, with
only a gradual
expansion.
At this point we have increased T
up to 500 $^\circ$K. Here  another fast
transition occurred as is clearly shown by the behaviour
of cell shape, as visible in the right  part of fig.
\ref{1}.
The $g(r)$ and $f(\theta)$ of this novel
structure are reported in fig. \ref{2}c.
Atoms are now again four-fold coordinated and the $g(r)$ is quite similar to
the starting $g(r)$ relative to the diamond structure. The width of the
peak in $f(\theta)$ is however quite large ($\sim 14^o$), and the final
structure is probably closer to  amorphous silicon than to a
proper crystal. Also, other features characteristic of the D-structure became
sizable (fig. \ref{1}c), but never quite as large as in the
perfect crystal. This amorphous phase is similar to those previously
obtained by first principles simulation via melt quenching \cite{ivan}.
In view of the well known metastability and slow relaxation rates of
covalent amorphous states, we were very satisfied with this result and
 did not attempt to anneal further to an ordered D phase.

Summarizing, the above example demonstrates that fast and efficient first
principles  fully dynamical simulations are now feasible, by merging
together  the {\sl ab-initio} CP molecular dynamics with the deformable
cell PR technique. The method appears to have the flexibility
required for a structural transformation to take place
between phases which are extremely
different, within a relatively short simulation time. The process of
transformation, and even the defects it inevitably implies, may themselves
be quite informative. We believe that it will thus  be possible
to uncover many
obscure parts of solid-state phase diagrams in the forthcoming future.

This work has been partially supported by the
{\sl Progetto Finalizzato Sistemi Informatici e Calcolo Parallelo},
by INFM, and by
the Research Office of the US Army.



\begin{figure}
 \caption{Time history of a compression and decompression molecular-dynamics
run on Si. Cell edges (a), their relative angles (b), and
the largest peaks in the ionic structure factor (c) are shown
as a function of time. The {\bf k} vectors of S({\bf k}) are labelled by the
Miller indices of the time-varying supercell.
The arrows denote the points where the external
pressure was
raised to the specified value.
T is 300 $^\circ$K\  up to 11 ps where it was raised to 500 $^\circ$K.}
 \label{1}
\end{figure}

\begin{figure}
 \caption{Pair correlation function $g(r)$ (left panels) and
angle distribution function $f(\theta)$ (right panels) at three
selected times during
the molecular-dynamics run described in the text.
  In a) and b) the arrows indicate the peaks position
expected for perfect D and sh crystals respectively.
$f(\theta)$ is computed including atoms
with coordination four in D and a-Si structures, and coordination eight in sh
structure.
$g(r)$ are plotted to $r$ larger than the half width of the
simulation cell
to show the effect of the cell shape on crystal structures.}
 \label{2}
\end{figure}

\begin{figure}
 \caption{Snapshot of an hexagonal plane of the defected sh structure
obtained in the simulation. Darker regions indicate defective sh.
A  hexagon in the perfect sh region is also shaded.
Several periodically repeated simulation cells are shown.}
 \label{3}
\end{figure}

\begin{figure}
 \caption{Evolution of the electronic states in the compression process.
Panel a) shows the
electronic states $\epsilon_i$ at the $\Gamma$ point of
the 54-atoms simulation cell as a function of time.
The two central panels show $\epsilon_i$
at $\Gamma$ of b) triclinically distorted and c) perfect sh 57-atoms cell
(see text).
Panel d) shows the electronic density of states  relative to the potential
of c).
Filled states are represented with continuous lines,
empty states with dash-dotted lines.
}
 \label{4}
\end{figure}


\begin{references}
\bibitem[17]{nota}
The relaxation of the triclinically distorted sh ends up in a quasi-perfect
sh (1.5 \% residual distortion), if the non-spherical cutoff imposed by our
constant plane waves technique is used. This residual distortion is
eliminated by imposing a spherical cutoff.
The $a$ and $c$ values given in the
test refer to this latter case.
\end{references}
\end{document}